----------------------------------------------------------------------------------------
\documentstyle[12pt]{article}

\def\Fig{{\bf Fig.}}
\def\S{{\bf S}}
\def\T{{\bf \Theta}}
\def\P{{\bf \Pi}}
\def\s{{\bf s}}
\def\H{{\bf H}_{qm}}
\def\U{{\bf U}_{T}}

\begin{document}
\begin{flushleft}
\vspace{-1 cm}
NEIP-95-003
\end{flushleft}
\begin{flushright}
\vspace{-1cm}
March 1995\\
\vspace{2 cm}
\end{flushright}
\begin{center}
\Large
{\bf  Evidence for the Validity of the Berry-Robnik Surmise in a Periodically
Pulsed Spin System} \footnotemark \footnotetext{Classification Numbers : 0545 /
0365}\\
\vspace{1 in}
\large
Ph. Jacquod \footnote{e-mail jack@iph.unine.ch}
 and J.-P. Amiet\footnote{e-mail amiet@iph.unine.ch}\\
\vspace{0.2in} Institut de Physique \\
Universit\'e de Neuch\^atel \\
1, Rue A.L. Breguet  \\
CH - 2000 Neuch\^atel \\
\end{center}
\normalsize
\vspace{0.4in}
\newpage
\begin{center}
{\bf Abstract }
\end{center}
\vspace{1 cm}
We study the statistical properties of the spectrum of a quantum dynamical
system whose classical counterpart has a mixed phase space structure consisting
of two regular regions separated by a chaotical one. We make use of a simple
symmetry of the system to separate the eigenstates of the time-evolution
operator into two classes in agreement with the Percival classification scheme
\cite{Per}. We then use a method firstly developed by Bohigas et. al.
\cite{BoUlTo} to evaluate the fractional measure of states belonging to the
regular class, and finally present the level spacings statistics for each
class. The level spacings distribution of states belonging to the irregular
part of the spectra as well as that of the complete set of levels corroborate
the Berry-Robnik surmise \cite{BeRo}. We further present a statistical study of
the regular  levels. The presence of intermediate states - states which belong
to neither class as long as $\hbar$ is finite, phase spatially mixed among the
set of regular ones, together with the small fractional measure of regular
states strongly affects the corresponding level spacings statistics, resulting
in a non negligible deviation from the expected Poisson distribution. We see
however the remarkable agreement of the irregular level spacings statistics as
a direct confirmation of the Berry-Robnik surmised.

\newpage
\section{Introduction}
\noindent For more than a decade, the study of quantum
mechanical systems whose classical counterpart exhibits chaos
has attracted much interest. One motivation for this study
is the paradoxical fact that while the correspondence principle,
as we understand it, should imply a quantum manifestation of
classical chaos, the Schr\"odinger equation is linear. As a
consequence, the time-evolution operator is unitary, and this
suppresses any exponential divergence in the time evolution of quantum states.
As a spectacular manifestation of this fact, time-reversal invariant models
show no lost of memory : reversing the time at a certain moment T brings us
back to the initial situation after another time interval T, while this would
require infinite precision in a classical chaotic system. Thus a basic
manifestation of classical chaos seems to have no place in quantum mechanics.

\noindent On the other hand, the destruction of an integral of motion,
of a quantum number, has striking effects on the statistical properties
of quantum spectras. It is today taken as granted that in a classically
integrable
system, the levels are uncorrelated, and so have a poissonian
level spacings distribution \cite{BeTa} (a remarkable exception being the
one-dimensional harmonic oscillator) and that in classically fully
chaotic models, the level spacings distribution has a dramatically different
shape :
it obeys predictions of random matrix theory, i.e. it exhibits level repulsion
\cite{Pech}.
\noindent The situation in mixed systems, where
regular and chaotic regions coexist in the classical phase space, is more
intricated. In an old paper Percival \cite{Per} classified the
eigenfunctions of the Schr\"odinger equation into two classes belonging
to either the regular regions, where the invariant tori are not destroyed,
or the chaotic one. This classification was based mostly on the correspondence
principle and has been numerically confirmed a few years ago
by Bohigas et. al. \cite{BoUlTo}. While the eigenfunctions that are mostly
confined on classically regular regions - we will call them the regular
eigenfunctions -
tend to concentrate on invariant tori, the irregular ones tend to spread
uniformly over
the chaotic region as $\hbar \rightarrow 0$, as has been rigorously
demonstrated by
Shnirelman \cite{Shni}. This picture is assumed to reflect reality in the
semiclassical limit $\hbar = 0$. Following this classification,
Berry and Robnik postulated that the part of the spectrum that corresponds
to regular eigenfunctions, has a poissonian level spacings distribution in
opposition to
the one corresponding to the irregular eigenstates which exhibits level
repulsion \cite{BeRo}. This surmise led them to an expression for the
level spacings distribution for mixed systems that has been
observed convincingly only recently \cite{ProRo1} for the case of the kicked
rotator on a torus. As pointed out by Prosen \& Robnik and Li \& Robnik
\cite{ProRo} \cite{RoBao}, reasons for this difficulty of observation
could be that we are not deep enough in the semi-classical regime. As long as
$\hbar$
is finite, a certain number of wave-functions belong neither to the regular nor
to the irregular set of eigenfunctions. We may think of states making use of
the Heisenberg uncertainty to overlap the frontier between the regular and
irregular regions of the classical phase space, or states located on the
regular region which, due to the finiteness of the Planck constant, do not yet
belong to the set of regular states. Consequently, the Berry-Robnik regime
should be observable only in the far semiclassical limit. We will come back to
this point later.

\noindent In this paper we present a spin model allowing a precise study of a
mixed regime. The reasons for this are first that, in a special regime, an
approximate simple symmetry of the phase space structure, namely $S_{z}
\rightarrow -S_{z}$, allows the separation of regular
states from the irregular ones, and secondly that the frontier between the
regular and the chaotic zones is rather sharp, thus minimizing the number
of intermediate eigenstates. This enables us to compute the level spacings
statistics
independently for the regular and irregular states. We interpret the fact that
these statistics obey quite well the poissonian distribution and the GOE
respectively as
a direct confirmation of the Berry-Robnik surmise.

\noindent We study the quantum system defined by the following Hamiltonian
\footnotemark \footnotetext{We use bold characters in the quantum case in
contrast to normal ones which refer to the classical and semiclassical cases :
${\vec \S }$ refers to the quantum spin operator while ${\vec S}$ is either a
classical or a semiclassical spin.}:
\begin{eqnarray}
\H := \frac{m}{2} ((1-z^2) \S_{z}^{2} - z^2 \S_{x}^{2}) + \kappa
\S_{z} \Delta_{T}
\end{eqnarray}
and the corresponding unitary time evolution operator :
\begin{eqnarray}
\U:= \exp(-\frac{i}{\hbar}\kappa \S_{z}) \exp(-\frac{i}{\hbar} \frac{m}{2}
((1-z^2) \S_{z}^{2} - z^2 \S_{x}^{2}) T)
\end{eqnarray}
where $\vec{\S} = (\S_{x},\S_{y},\S_{z}) = \hbar  (\s_{x},\s_{y},\s_{z}) =
\hbar \vec{\s}$ are spin operators satisfying the usual commutation rules
($\epsilon^{ijk}$ is the total antisymmetric tensor of 3$^{rd}$ order):
\begin{eqnarray}
\left[ \S_{i},\S_{j} \right] = i \hbar \epsilon^{ijk} \S_{k}
\end{eqnarray}
$0 \leq \kappa\leq 2 \pi $,
$ \Delta_{T}:=\sum_{n=-\infty}^{+\infty} \delta(t-n T)$,   [$m$] =
energy$^{-1}$ time$^{-2}$ and $0\leq z \leq 1$ . Models of this kind have been
extensively studied \cite{haa}. They represent a spin which evolves under the
influence of a classically integrable Hamiltonian $\H^{0} = \frac{m}{2}
((1-z^2) \S_{z}^{2} - z^2 \S_{x}^{2})$  during a time $T$ after which the spin
undergoes a rotation of angle $\kappa$ around the z-axis.
\noindent The regime we consider is defined by $\kappa = 1.1$, $T=\frac{19}{m
S}$
and z$^{2} = \frac{1}{2}$. Classically there are two regular zones around the
north
and south poles surrounding a chaotic region which is fairly well
symmetric under S$_{z}$ reflection (Fig.1). In the semiclassical limit which
corresponds to
$\hbar s = S = $ constant, $\hbar = s^{-1} \rightarrow 0$, states
which are located on the chaotic region tend to cover it homogeneously
according
to Shnirelman's theorem \cite{Shni}. Since this region is symmetric under
S$_{z}$ reflection, the expectation value $<\Psi_{chaos}|\s_{z}|\Psi_{chaos}>$
of such a state
tends to disappear as we approach the semi-classical limit. For small but
finite $\hbar$, the distribution of  $<\Psi_{k}|\s_{z}|\Psi_{k}>$, where
$|\Psi_{k}>$ is an eigenstate of the operator $\U$ defined in (2), will then
present a sharp peak around zero corresponding to
the irregular states surrounded by two smaller bumps corresponding to regular
states (Fig.2). This allows us to separate easily the regular states from the
irregular ones, the validity of this selection being confirmed by a numerical
semiclassical argument presented in section 3 as well as an extensive study of
the Husimi distributions of the selected states \cite{AmJa}.

\noindent The paper is organized as follows : Section 2 is devoted to a short
presentation of the classical model. In section 3 we derive some useful
semiclassical quantities such as the density of states and the expression for
the action. This will allow us to estimate the number of regular states, and
give a check of our selection criterion. In section 4 we present the quantum
mechanical
model as well as our numerical results for a spin magnitude $s$=500. All of
them were obtained using
direct diagonalization techniques. Conclusions and further remarks are given in
section 5.
\section{Classical model}
The unperturbed classical Hamiltonian
\begin{eqnarray}
H_{cl}^{0} := \frac{m}{2} ((1-z^2) S_{z}^{2} - z^2 S_{x}^{2})
\end{eqnarray}
has two degrees of freedom and is an integral of motion. The trajectories are
confined to the intersections of the sphere $|\vec{S}| = S$ with the cones of
constant energy
$E = \frac{m}{2} ((1-z^2) S_{z}^{2} - z^2 S_{x}^{2})$. The perturbation :
\begin{eqnarray}
H_{cl}^{1} := \kappa S_{z} \Delta_{T}
\end{eqnarray}
corresponds to a rotation of angle $\kappa$ around the z-axis performed at time
intervals $T$. Its addition
leads to the destruction of the energy surfaces, and allows more and more
trajectories
to wander chaotically on the sphere of constant spin magnitude as $\kappa$ and
$T$ grow. Expanding $H_{cl}^{0}$ up to the first order in $\delta
S_{z}:=S-S_{z}$ near the poles $S_{z} = \pm S$,we get a one-dimensional
harmonic oscillator of period $T=2 \sqrt{2} \frac{\pi}{m S}$. In particular we
have $\dot{\delta S_{z}}=O(\delta S_{z}^{2})$ : in this approximation $\delta
S_{z}$ is an integral of motion and is furthermore conserved by the
perturbation $H_{cl}^{1}$ too. It is thus conceivable that the invariant torii
near the poles will offer more resistance to the perturbation than those
located away from them. We use this property to find a regime in which there
are two regular islands around the poles approximately related by the operation
$S_{z} \rightarrow -S_{z}$ and separated
by a chaotic region. This we achieved by setting  $\kappa = 1.1, T=\frac{19}{m
S}, z^{2}=0.5$  (Fig.1). The regular islands occupy in a good approximation
the region $ 0.22 S^{2} \leq E \leq E_{max} = 0.25 S^{2} $.
\section{Semiclassical approach}
We compute the Green function for a trajectory of positive energy and the
density of states for the unperturbed case $T=\frac{19}{m S}, z^{2}=0.5$. We
follow the lines drawn in \cite{Rei}. We first write the unperturbed
Hamiltonian in canonical variables (S$_{z},\phi$) for the chosen regime :
\begin{eqnarray}
H_{0} = \frac{m}{4} (S_{z}^{2} (1+\cos^{2}(\phi)) - \vec{S}^{2} \cos^{2}(\phi))
\end{eqnarray}
The action integral for a trajectory of energy $E$ starting at $\phi_{0}$ and
ending at $\phi$ reads ($\vec{S} = \hbar \vec{s}$, $e=\frac{E}{\hbar^{2} m}$):
\begin{eqnarray}
{\cal S}_{\beta}(\phi_{0},\phi,e) =
\hbar \left[\int_{\phi_{0}}^{\phi^{*}} \sqrt{\frac{4 e+s^{2}
\cos^{2}(\phi')}{1+\cos^{2}(\phi')}} d\phi'+
n_{\beta} \int_{0}^{2 \pi} \sqrt{\frac{4 e+s^{2}
\cos^{2}(\phi')}{1+\cos^{2}(\phi')}} d\phi'\right]
\end{eqnarray}
where we have set $\phi^{*} = \phi_{0}  +  (\phi-\phi_{0}) $ mod $2 \pi$, and
$n_{\beta}$ is the number of complete revolutions
accomplished between $\phi_{0}$ and $\phi$ ($\phi= 2 \pi n_{\beta} + \phi*$).
The sum runs over classical orbits $\beta$ of constant energy.
This leads us to the expression for the corresponding Green function :
\begin{eqnarray}
{\cal G}(\phi_{0},\phi,e) & = & -\frac{i}{\hbar} \sum_{\beta} \sqrt{|{\det \cal
D}_{1,\beta}(\phi_{0},\phi)|} \exp(\frac{i}{\hbar}
{\cal S}_{\beta}(\phi_{0},\phi,e)-\frac{i \pi}{2} l_{\beta}) \nonumber \\
& =: & \sum_{n} \frac{\psi_{n}^{*}(\phi) \psi_{n}(\phi_{0})}{(E-E_{n}+i
\epsilon)}
\end{eqnarray}
with
\begin{eqnarray}
& & \hspace{-2cm} \det {\cal D}_{1,\beta} (\phi_{0},\phi) =
\frac{1}{m^{2}}\left({\partial^{2}{{\cal S}_{\beta}} \over \partial \phi
\partial \phi_{0}} {\partial^{2}{{\cal S}_{\beta}} \over \partial e^{2}} -
{\partial^{2}{{\cal S}_{\beta}} \over \partial \phi \partial
e}{\partial^{2}{{\cal S}_{\beta}} \over \partial \phi_{0} \partial e} \right) =
\nonumber \\
& & \hspace{-2cm} \frac{-4}{m^{2} \sqrt{(4 e+s^{2} \cos^{2}(\phi_{0}))
(1+\cos^{2}(\phi_{0})) (4 e+s^{2} \cos^{2}(\phi)) (1+\cos^{2}(\phi))}}
\end{eqnarray}
this result being obtained by partial differentiations of (7). Since this
latter value never changes sign, the Maslov index  $l_{\beta}$ vanishes and so
the divergence of the Green function leads to the following semiclassical
quantization condition :
\begin{eqnarray}
\int_{0}^{2 \pi} \sqrt{\frac{4 e+s^{2} \cos^{2}(\phi')}{1+\cos^{2}(\phi')}}
d\phi' = 2 \pi M
\end{eqnarray}
for any integer $0 \leq M \leq s$. We have then for the averaged density of
states (N is the number of states) :
\begin{eqnarray}
 \overline{\rho}(e) & = & -\frac{1}{N \pi} Im\left[ \int_{0}^{2 \pi} d\phi
{\cal G} (\phi,\phi,e) \right] \nonumber\\
{} & = &  \frac{1}{\pi \hbar} \int_{0}^{2 \pi} d\phi \sqrt{|{\det \cal
D}_{1,\beta}(\phi,\phi)|}\nonumber \\
{} & = & \frac{2}{\pi \hbar m} \int_{0}^{2 \pi} \frac{d\phi}{\sqrt{
(4 e+s^{2} \cos^{2}(\phi)) (1+\cos^{2}(\phi)))}}
\end{eqnarray}
The last equation states in particular that the averaged density of states is
proportional to the classical orbit period. Fig. 3 shows the agreement of this
semiclassical result with the numerically obtained density of states for the
unperturbed quantum model at $s$=1000.
Using (11) we estimate the number of states occupying the regular region of
figure 1 :
\begin{eqnarray}
N_{reg} \approx \int_{0.22 s^{2}}^{0.25 s^{2}} \overline{\rho}(e) de\approx
\frac{1}{24} (2 s+1)
\end{eqnarray}
The number of states occupying this region in absence of perturbation is
$\frac{1}{24}$ times the total number of states. This gives us a first
approximation for the number of regular states we must select. A better
approximation in presence of perturbation is given using a method developed by
Bohigas et. al. \cite{BoUlTo}. We must evaluate the number
N$_{reg}$ of trajectories that satisfy the condition :
\begin{eqnarray}
\sum_{i=0}^{P} \left[ \int_{\phi_{i}^{+}}^{\phi_{i}^{-}} s_{z}(\phi') d\phi'
+ s_{z}(\phi_{i}^{+}) \kappa \right] = 2 \pi M
\end{eqnarray}
for some integers $M$ and $P$, while it nearly closes on itself after the
$P^{th}$ kick, i.e. : $ s_{z}(\phi_{M}^{+}) \approx s_{z}(\phi_{0}^{+})$.
$\phi_{i}^{-}$ is the angle between the x and the y component of the
spin just before the i$^{th}$ kick while $\phi_{i}^{+}$ refers to the same
angle right after this kick.

\noindent This condition means that the action integral must still be
an integer multiple of 2 $\pi$, and that simultaneously, the orbit must be
closed. This condition
has meaning only on regular regions were the invariant torii are not destroyed,
so that the
integrals make sense. We transform this condition and compute the number of
trajectories satisfying :
\begin{eqnarray}
\frac{\sum_{i=0}^{P} \left[ \int_{\phi_{i}^{+}}^{\phi_{i+1}^{-}} s_{z}(\phi')
d\phi'
+ s_{z}(\phi_{i}^{+}) \kappa \right]}{\kappa P + \sum_{i=1}^{P} \left(
\phi_{i}^{-} - \phi_{i-1}^{+} \right)} \approx M
\end{eqnarray}
for integers $M$ and $P$, and $P$ sufficiently large. With this we replace two
conditions by only one numerically more tractable condition. Since our task is
to evaluate the number of regular semiclassical levels, and not to determine
them precisely, we believe that condition (14) is sufficient. The number of
regular states  we numerically estimated with (14) is 50 $\pm$ 4 for $s$=500,
i.e. slightly larger than that estimated with (12). In the next section, we
will consider this estimated number of regular states as a check of the
validity of our selection criterion.
\section{Quantum Model}
In this section we study the statistical properties of the spectrum of the
quantum Hamiltonian (1) for integer spin magnitude.  Since the perturbation
term is time-dependent, the energy is no longer a good quantum number, and we
are led to define quasi-energies and quasi-energy eigenstates. The
Schr\"odinger equation leads to the following time evolution from right after a
kick to right after the next one:
\begin{eqnarray}
\Psi (T^{+}) = \U \Psi(0^{+}) =
 \exp(-\frac{i}{\hbar} \kappa \S_{z}) \exp(-\frac{i}{\hbar} \H^{0} T)
\Psi(0^{+})
\end{eqnarray}
Quasi-energies $\lambda$ and quasi-energy eigenstates $\Psi_{\lambda}$ are then
defined by :
\begin{eqnarray}
\U \Psi_{\lambda} = \exp(-i \lambda) \Psi_{\lambda}
\end{eqnarray}
Since $\U$ is unitary, the $\lambda$'s are real and defined modulo 2 $\pi$. We
introduce two parities :
\begin{eqnarray}
\P | \mu > = |- \mu> \\
\T |\mu> = (-1)^{s-\mu} |\mu>
\end{eqnarray}
We can express the time-reversal operator {\bf T} in term of these two parity
operators :
\begin{eqnarray}
\P \circ \T  | \mu> = {\bf T} | \mu > = (-1)^{s-\mu}  |- \mu>
\end{eqnarray}
 In the integer spin case the eigenstates $|\Psi>$ of $\U^{0} :=
\exp(-\frac{i}{\hbar} \H^{0} T) $ satisfy the conditions :
\begin{eqnarray}
\P |\Psi> = \pm |\Psi> \\
\T |\Psi> = \pm |\Psi>
\end{eqnarray}
So $ \H^{0} $ and $ \U^{0} $ are in particular time-reversible. The
perturbation breaks the $\P$-parity but leaves the $\T$-parity unbroken
\footnotemark \footnotetext{In the half-integer spin case, eigenstates of the
unperturbed time-evolution operator are eigenstates of the $\T$-parity only.
The latter is left unbroken by the perturbation we consider.}. We will
concentrate on the study of even states, i.e. those states satisfying :
\begin{eqnarray}
\T |\Psi> =  |\Psi>
\end{eqnarray}
\noindent However partial results obtained for the odd set of states
corroborate the results presented here. The key point is now to find a clear
quantum manifestation of the approximate symmetry $S_{z} \rightarrow -S_{z}$ of
the classical phase space structure. A practical solution is given by
Shnirelman's theorem which states that in the semiclassical limit, the quantum
states that are confined on the classically chaotic region of the phase space
tend to cover it uniformly. To get an insight in this statement we use the
following resolution of unity \cite{Pelo}
\begin{equation}
{\bf 1} = \frac{2 s+1}{\pi} \int d\theta d\phi \sin \theta
|\theta,\phi><\theta,\phi|
\end{equation}
where we introduced coherent states of the spin $SU(2)$ group :
$$  |\theta,\phi> := \sum_{\mu=-s}^{s}\sqrt{\left(^{\hspace{0.15cm}2 s}_{s-\mu}
\right)} \sin(\frac{\theta}{2})^{s-\mu} \cos(\frac{\theta}{2})^{s+\mu} e^{i
(s-\mu) \phi} |\mu > $$
These are states that are centered on the point ($\theta,\phi$) of the sphere
and which minimalize the quantum uncertainty. $\theta $ is defined by $S_{z} =
S \cos(\theta)$. Using (22), the symmetry of the chaotical region and
Shnirelman's theorem \cite{Shni}  :
\begin{equation}
<\Psi_{chaos}|\theta,\phi>  \longrightarrow \cases{
0 & on the regular region\cr
const & on the chaotic region
}
\end{equation}
it is then easy to show that
\begin{equation}
<\Psi_{chaos}|\s_{z}|\Psi_{chaos}> \rightarrow 0
\end{equation}
in the semiclassical limit. This translates into Fig.2 where we plotted an
histogram of the expectation value of $\s_{z}$ taken over quasi-energy
eigenfunctions for $s$=500, $\kappa=1.1$, z$^{2}$=0.5, and $m$=1. The central
peak clearly reflects our reasoning, while the two smaller bumps surrounding it
are mainly due to the regular states that are confined to the classical
stability islands. The gap in-between is a consequence of the uniform
distribution of irregular states. It is remarkable that this gap overlaps the
classical frontier between regular and chaotic region.

\noindent We used this property to part the irregular states from the regular
ones and then study separately the statistical properties of the spectrums of
each class of states. We believe this criterion is justified since the
fluctuations
\begin{eqnarray}
 \Delta \s_{z} = \sqrt{<\s_{z}^{2}>-<\s_{z}>^{2}}
\end{eqnarray} of regular states is much smaller than the "Shnirelman gap"
appearing in the histogram of Fig. 2 between the huge central peak and the
smaller bumps. As a consequence only very few regular levels will be selected
with the set of irregular ones, while maybe more irregular will be counted with
the regular ones. Moreover, the fact that the number of selected regular states
is in complete agreement with the numerical semiclassical evaluation given by
(14) confirms the relevance of this selection criterion.

\noindent We now turn our attention to the study of the spectral properties of
the time-evolution operator (16). Due to the $\T$-symmetry (18), $\U$ belongs
to the circular orthogonal ensemble, and not to the circular unitary ensemble
as would be expected from the fact that the perturbation breaks the
time-reversal symmetry. This situation is similar to the one encountered by
Berry \& Robnik in certain Aharonov-Bohm billiards \cite{RoBe}, or by Delande
\& Gay in the Hydrogen atom in a magnetic field \cite{DelGay} where the system
violates the time-reversal symmetry, but possesses an invariance under a
combination of the time-reversal and another symmetry, in our case the
$\P$-symmetry. We thus expect a linear repulsion for the part of the spectrum
belonging to the irregular states.

\noindent The results of our study for a spin magnitude $s$=500 are plotted in
Fig. 4 to 9. Fig. 4 shows a plot of the level spacings statistics for 4233
irregular level spacings computed by diagonalizing ten different evolution
matrices for $T=\frac{19}{m S} $ and $ 1.05 \leq \kappa \leq 1.15$. The solid
line is the predicted Wigner distribution. The agreement is excellent. In Fig.
5 we plotted the corresponding cumulative level spacings distribution defined
in term of the level spacings distribution \footnotemark \footnotetext{We have
used $\underline{s}$ for the level spacings to avoid confusion with the spin
magnitude.} $P(\underline{s})$ by
\begin{eqnarray}
W(\underline{s}) = \int_{0}^{\underline{s}}dt P(t)
\end{eqnarray}
Also shown are the Poisson and the Wigner distributions. As shown in inset,
small deviations from the Wigner distribution appear only around spacings
$\underline{s} \approx$2, but have no significance to our opinion.

\noindent Fig.6 and 7 show the level spacings and the cumulative level spacings
distribution for 572 regular levels taken from twenty different evolution
matrices for $T=\frac{19}{m S} $ and $ 1.05 \leq \kappa \leq 1.15$. The
difficulty here is the relatively small number of regular states ($\approx$
25). Accordingly, only few intermediate or irregular states can have relatively
big effects on the statistics. In a convenient basis, $\U$ can be represented
by the following matrix :
\begin{eqnarray}
 \U := \left( \begin{array}{c c} R & K \\ K & C \end{array} \right)
\end{eqnarray}
$R = R_{i} \delta_{i,j}$ is a diagonal N$_{reg} \times$ N$_{reg}$ matrix
which corresponds to the N$_{reg}$ regular states, $C = C_{i} \delta_{i,j}$ is
a diagonal
N$_{chaos} \times$ N$_{chaos}$ matrix which corresponds to the N$_{chaos}$
irregular states,
and $K = {\rm O}(\hbar)$ couples the two subspaces as long as $\hbar$ is
finite. In our picture $K$ disappears in the semiclassical limit, and the
$R_{i}$ and $C_{i}$ satisfy
a poissonian statistics and a GOE statistics respectively. It would be of
course a hopeless task to try to determine K for finite $\hbar$. The important
point is to recognize that as long as $\hbar$ is finite but small enough, $K$
couples only few regular states with irregular ones, this fact resulting in a
deviation from the Berry-Robnik surmise. This deviation is then naturally much
more important for the regular part of the spectrum, since it contains much
fewer levels than the
irregular part. We believe that this is the reason for the deviation of the
statistics of the set of levels we have selected as regular from the poissonian
predicted behaviour. We must recall that our whole reasoning is
based on the assumption of two classically homogeneous stability islands. In
such a case,
semiclassical wave-function would mimic classical orbits and would therefore
fit together as
concentric circles. The presence of hyperbolic fixed points or cantori may
change this picture, possibly turning regular states into intermediate ones as
long as $\hbar$ is finite. The semiclassical wave-function overlap and thus
interact at certain regions, and this, in the Pechukas picture \cite{Pech},
modify very sensibly the equations of motions governing the evolution of the
quasi-energies $\lambda$ as $\kappa$ or $T$ is modified, resulting in the
appearance of level repulsion. So some intermediate states are phase spatially
mixed among the set of states we have selected as regular and consequently
modify the corresponding statistics. Their effect is furthermore enhanced by
the small ratio of regular levels. A current investigation of the Husimi
densities of the selected regular states corroborates this reasoning
\cite{AmJa}.
\noindent Finally we show in Fig.8 and 9 level spacings and cumulative level
spacings statistics for the complete set of levels. We compare our results with
the Berry-Robnik prediction for a fractional measure of regular states as
approximated by (12). The agreement is amazing, and corroborates our picture.
The $\chi^{2}$-test for both graphs - $\chi^{2}$=25, i.e. half the number of
boxes for Fig.8, and $\chi^{2}$=1480, i.e. 3.3 times less than the number of
levels for Fig.9 - gives full statistical significance to these last graphs. We
see them as a good evidence for the validity of the Berry-Robnik surmise in our
model.

\section{Conclusion}
We studied the statistical properties of a quantum spin model whose classical
counterpart exhibits a mixed phase space configuration. Due to a simple
approximate symmetry, whose effect on the quantum system is drastically
enhanced by Shnirelman's theorem, we were able to separate the irregular from
the regular levels, thereby confirming implicitely the validity of the Percival
classification.  We then performed a separated statistical study of these
levels. The results confirm the Berry-Robnik surmise : while the irregular set
of quasienergies exhibits a clear wigner-like shape, the regular part of the
spectrum has a clearly different shape, though its spacings distribution does
not follow strictly a poissonian law. This deviation is interpreted as the
presence of both irregular and intermediate states among the selected regular
ones, their effect being enhanced by the relatively small number of the latter.
Nevertheless, due to the small number of regular states we believe that the
irregular statistics is much more significant, and see our results as a good
confirmation of the validity of the Berry-Robnik surmise in our model.
\\ \\ \\
\Large
\noindent \bf{Acknowledgements} \\ \\
\normalsize
One of us (P.J.) gratefully acknowledges fruitfull discussions with J.
Bellissard, C. Rouvinez and D. Shepelyansky , as well as the hospitality of the
theoretical physics division of the "Laboratoire de Physique Quantique,
Universit\'e Paul Sabatier" in Toulouse extended to him during his visit when
part of this work has been done. We are grateful to T. Prosen for having drawn
our attention on reference \cite{ProRo1}. Work supported by the Swiss National
Science Foundation.

\newpage
\section*{Figure Captions}
\Fig {\bf 1}: Orthogonal projection of the classical phase space on the
($S_{x}$,$S_{y}$) plane for the case T=$\frac{19}{m S}$, $\kappa=1.1$ and
z$^{2}$=0.5. \\ \\
\Fig {\bf 2}: Histogram of the expectation value of $S_{z}$ taken over the
eigenstates of the unitary time evolution operator defined in (2). \\ \\
\Fig {\bf 3}: Density of states for the unperturbed Hamiltonian according to
(11) (solid line) as compared to numerically obtained datas for the case
$s$=1000 (squares). \\ \\
\Fig {\bf 4}: Level spacings distribution for 4233 irregular level spacings
obtained through direct diagonalization of ten evolution matrices in the
parameter range $T=\frac{19}{m S}$ and
$1.05 \leq \kappa \leq 1.15$. \\ \\
\Fig {\bf 5}: Cumulative level spacings distribution for the same case as
fig.4.  In inset : regions of small deviation relatively to the
Wigner-distribution. \\ \\
\Fig{\bf 6}: Level spacings distribution for a set of 472 regular level
spacings obtained through direct diagonalization of twenty evolution matrices
in the parameter range $T=\frac{19}{m S}$ and
$1.095 \leq \kappa \leq 1.105$. The solid line is the predicted Poisson
distribution.\\ \\
\Fig{\bf 7}: Cumulative level spacings distribution for the same levels as
Fig.6 compared to the Poisson distribution. \\ \\
\Fig{\bf 8}: Level spacings distribution for a set of 5000 regular and
irregular level spacings obtained through direct diagonalization of ten
evolution matrices in the parameter range $T=\frac{19}{m S}$ and
$1.095 \leq \kappa \leq 1.105$. The solid line is the predicted Berry-Robnik
distribution with
fractional measure of regular states $\rho_{1} = 0.08$. $\chi^{2}$=25 is half
the number of boxes.\\ \\
\Fig{\bf 9}: Cumulative level spacings distribution for the same levels as
Fig.8 compared to the poissonian and the Berry-Robnik predicted distribution.
In inset : Same curve compared to the
Wigner distribution. $\chi^{2}$=1480 is 3.3 times less than the number of
levels.
\end{document}